\documentclass[prd,superscriptaddress,a4paper,showpacs,showkeys,11pt,nofootinbib]{revtex4}
\usepackage{graphicx} 
\usepackage{graphicx}
\usepackage{amsmath}
\usepackage{booktabs}
\usepackage{siunitx}
\usepackage{url}
\usepackage{hyperref}
\begin{document}
\title{Investigating the exclusive toponium production at the LHC and FCC}

\author{Reinaldo {\sc Francener}}
\email{reinaldofrancener@gmail.com}
\affiliation{Instituto de Física Gleb Wataghin - Universidade Estadual de Campinas (UNICAMP), \\ 13083-859, Campinas, SP, Brazil. }

\author{Victor P. {\sc Gon\c{c}alves}}
\email{barros@ufpel.edu.br}
\affiliation{Institute of Physics and Mathematics, Federal University of Pelotas (UFPel), \\
  Postal Code 354,  96010-900, Pelotas, RS, Brazil}

\author{Daniel E. {\sc Martins}}
\email{daniel.ernani@ifj.edu.pl}
\affiliation{The Henryk Niewodniczanski Institute of Nuclear Physics (IFJ)\\ Polish Academy of Sciences (PAN), 31-342, Krakow, Poland
}

\begin{abstract}
An exploratory study of the exclusive toponium production in $pp$, $pPb$ and $PbPb$ collisions at the center - of - mass energies of the Large Hadron Collider (LHC) and Future Circular Collider (FCC) is performed. Assuming that the toponium is a pseudoscalar $t\bar{t}$  state, we consider its exclusive production by photon  and gluon - induced interactions. Results for the total cross - sections and associated rapidity distributions are presented, and  the number of events  at the LHC and FCC are estimated.
\end{abstract}


\maketitle

\section{Introduction}

\begin{figure}[t]
	\centering
	\begin{tabular}{ccc}
    \\\includegraphics[width=0.45\textwidth]{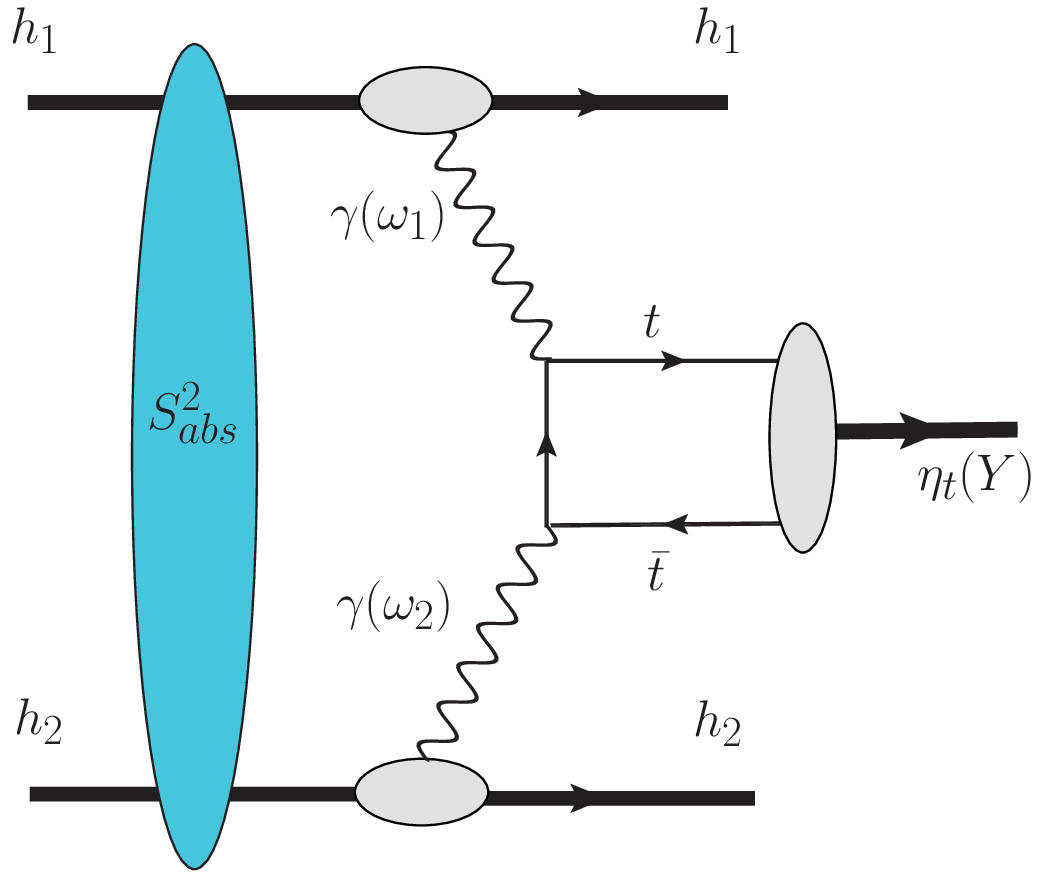} & \,&
    \includegraphics[width=0.5\textwidth]{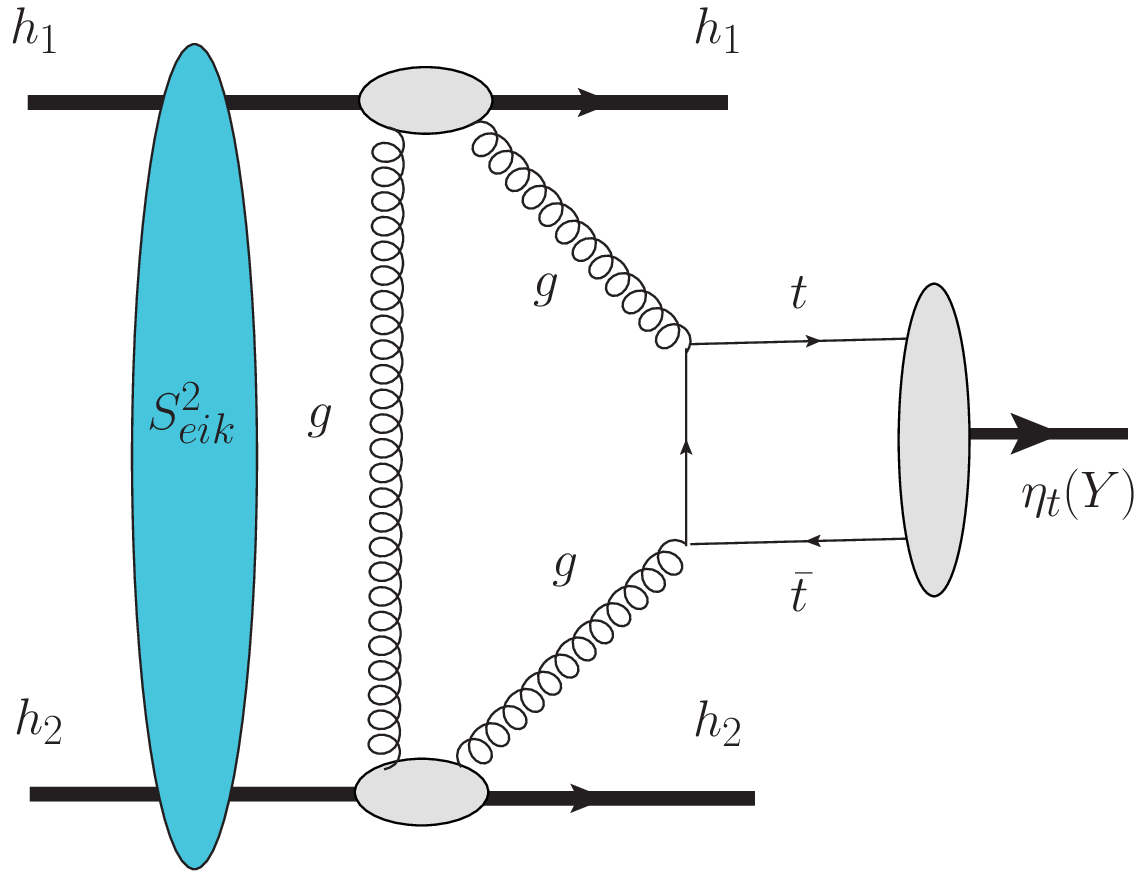}     
			\end{tabular}
\caption{Exclusive $\eta_t$ production by photon (left panel) and gluon - induced (right panel) interactions in hadronic collisions.}
\label{fig:diagram}
\end{figure}

The description of  bound
states of heavy quarks,  generically denoted as quarkonium,  is one of the main challenges  
of quantum chromodynamics (QCD). Over the last decades, a vast literature dedicated to the theoretical treatment and experimental observation of $c\bar{c}$ and $b\bar{b}$ bound states has been produced (For a review see, e.g.   Ref. \cite{Brambilla:2010cs}). In contrast, until recently, the discussion about the production in hadronic collisions of a color singlet bound state formed  by a top and an antitop quark - the toponium, was restricted to some few studies (See, e.g. Refs. \cite{Fadin:1990wx,Fabiano:1993vx,Hagiwara:2008df,Sumino:2010bv,Fuks:2021xje}).  Although the formation of toponium in $e^-e^+$ collisions had been predicted by Fadin and Khoze many years ago \cite{Fadin:1987wz}, the discovery of the top quark with a large mass and rapid decay have reduced the interest about the associated bound state, since it has been widely accepted that there is an insufficient time for the  bound state formation before top decay. However, the observation  of a 3$\sigma$ excess of dileptonic top-antitop events in the smallest relative angle bin by the ATLAS collaboration \cite{ATLAS:2019hau} and additional features in the high  statistic $t\bar{t}$ events observed by the ATLAS and CMS collaborations \cite{ATLAS:2023gsl,CMS:2024ybg}, have largely motivated several groups to revisit the  toponium production in hadronic collisions \cite{Aguilar-Saavedra:2024mnm,Fuks:2024yjj,Garzelli:2024uhe} as well as the derivation of the { different toponium states} \cite{Wang:2024hzd,Fu:2024bki,Jiang:2024fyw}. { In particular, the CMS collaboration has recently released the results associated with the search for resonances in $t\bar{t}$ production \cite{CMS:2025kzt}, where a significant excess of events is observed near the kinematical threshold. Although these results are consistent  with the formation of a pseudoscalar $t\bar{t}$ state, denoted $\eta_t$ state,  the observation (or not) of a toponium at the LHC is still a theme of debate. }

 The existing literature for the toponium production in hadronic collisions is focused on the  inclusive case, characterized by the fragmentation of the incident protons
(See, e.g. Refs.\cite{Aguilar-Saavedra:2024mnm,Fuks:2024yjj,Garzelli:2024uhe}). However, in recent years, the study of particle production in exclusive processes, where the incident hadrons remain intact in the final state, became a reality (For recent reviews see, e.g., Refs. \cite{upc,review_lang}). Motivated by the results obtained in Refs.~\cite{Goncalves:2020saa,Martins:2022dfg}, where the exclusive $t\bar{t}$ production was  investigated, and by the fact that{  for high energies} the pseudoscalar mesons are dominantly produced by gluon - gluon and photon - photon interactions,  we will calculate in this paper, for the first time, the exclusive toponium production by photon and gluon - induced interactions in hadronic collisions at the LHC and FCC energies { using the formalism developed in Refs.~\cite{Khoze:2000cy,Khoze:2001xm,Khoze:2000jm}.} The dominant diagrams are represented 
in Fig.  \ref{fig:diagram}, and  can be written in the form
\begin{equation}
h_1 + h_2 \rightarrow h_1 \otimes {\eta_t} \otimes h_2,
\end{equation}
where $h_i$ is a proton or a nucleus  and  ${\eta_t}$ represents the pseudoscalar $t\bar{t}$  state. The basic characteristic of these processes is the presence of two rapidity gaps ($\otimes$) in the final state, which are regions devoid of hadronic activity, separating the toponium from the intact outgoing hadrons. Experimentally, these processes have a very clear signal in the absence of pile-up, with the presence of the { decaying products of the $\eta_t$ state} and no other hadronic activity seen in the central detector. Moreover, these events can be separated at the LHC by measuring the outgoing hadrons using the forward proton
detectors (FPD) such as the ATLAS Forward Proton detector (AFP)
\cite{Adamczyk:2015cjy,Tasevsky:2015xya} and Precision Proton Spectrometer (CT-PPS) \cite{Albrow:2014lrm} that are installed symmetrically around the interaction point at a distance of roughly 210~m from the interaction point. Similar forward detectors are  expected to also be installed at the FCC \cite{FCC:2018vvp}. In what follows, we will present a brief review of the formalism needed to estimate both contributions for the exclusive $\eta_t$ production in hadronic collisions. For  detailed reviews { about the production of other final states in  exclusive processes}, we refer the interested reader to the Refs. \cite{upc,review_lang}. In addition, we indicate Refs. \cite{Goncalves:2020saa,Martins:2022dfg} for the related discussion about the exclusive open $t\bar{t}$ production in $pp$ collisions.

{ This paper is organized as follows. In the next section, we will present a brief review of the formalism used to describe the $\eta_t$ production by $\gamma \gamma$ and $gg$ interactions in exclusive hadronic processes. Moreover, we will discuss the main ingredients used in our calculations.   We will focus on the lowest $S$ - wave state { ($1S$ state)}, but the formalism can be directly applied for excited states { (e.g. $2S$ and $3S$ states)} if the corresponding masses and width decays are used in the calculations.  In section \ref{sec:res} we will present our predictions for the total cross - sections considering $pp$, $pPb$ and $PbPb$ collisions at the LHC and FCC energies, which demonstrate that the $\eta_t$ production is dominated by $\gamma \gamma$ interactions. In addition, the corresponding invariant mass and rapidity distributions will also be presented. Finally, in section \ref{sec:sum}, we will summarize our main results and conclusions.}

\section{Formalism}

The exclusive toponium production by $\gamma \gamma$ interactions in hadronic collisions at a center - of - mass energy $\sqrt{s}$, represented in Fig. \ref{fig:diagram} (left panel),  is given by \cite{baur_jpg}
\begin{eqnarray}
\sigma (h_1 h_2 \xrightarrow{\gamma \gamma} h_1 \otimes {\eta_t} \otimes h_2) =  \int_{0}^{\infty}\! \frac{d\omega_{1}}{\omega_{1}}\! \int_{0}^{\infty}\! \frac{d\omega_{2}}{\omega_{2}}\ F(\omega_1, \omega_{2})\ \hat{\sigma}_{\gamma \gamma \rightarrow {\eta_t}}(\omega_{1}, \omega_{2})
\end{eqnarray}
where $\hat{\sigma}_{\gamma \gamma \rightarrow {\eta_t}}$ is the cross-section for the subprocess $\gamma \gamma \rightarrow {\eta_t}$  and $\omega_1$ and $\omega_2$ are the energies of the photons which participate of the hard process and that can be  expressed in terms of photon - photon center - of - mass energy $W = 4 \omega_1 \omega_2  $ and rapidity $Y$ of the $\eta_t$ state  as follows: 
\begin{eqnarray}
\omega_1 = \frac{W}{2} e^Y \,\,\,\,\mbox{and}\,\,\,\,\omega_2 = \frac{W}{2} e^{-Y} \,\,\,.
\label{ome}
\end{eqnarray}
Moreover,  $F$ is the folded spectra of the incoming particles (which corresponds to an ``effective luminosity'' of photons) given by~\cite{baur_ferreira}
\begin{equation}
 F(\omega_1, \omega_{2}) = 2\pi \int_{0}^{\infty} db_{1} b_{1} \int_{0}^{\infty} db_{2} b_{2} \int_{0}^{2\pi} d\phi\ N_{1}(\omega_{1}, b_{1}) N_{2}(\omega_{2}, b_{2}) \langle {\mathcal S}^2_{abs} \rangle \,\,,
 \label{efe}
\end{equation}
where $N(\omega_i, {\mathbf b}_i)$ is the equivalent  spectrum  
of photons with energy $\omega_i$ at a transverse distance ${\mathbf b}_i$  from the center of hadron, defined in the plane transverse to the trajectory, and $\phi$ is the angle between $\mathbf{b}_1$ and $\mathbf{b}_2$. The impact parameter of the collision $\mathbf{b}$ is related to $\mathbf{b}_1$ and $\mathbf{b}_2$ by $b^2 = b_1^2 + b_2^2 - 2b_1 b_ 2 \cos \phi$ (For a detailed discussion, see e.g. Ref.  \cite{Azevedo:2019fyz}). 
The quantity $\langle {\mathcal S}^2_{abs} \rangle$ in Eq. (\ref{efe}) ensures that the { incoming particles} do not overlap, which we assume to be given by  $\langle {\mathcal S}^2_{abs} \rangle = \Theta(b - R_{A} - R_{B})$  \cite{baur_ferreira}, where  $R_i$ represents the { proton or Lead radius}. The Weizs\"acker-Williams photon spectrum for a given impact parameter is given in terms of the nuclear charge form factor $F(k_{\perp}^2)$, where $k_{\perp}$ is the four-momentum of the quasi-real photon, as follows  \cite{upc}
\begin{eqnarray}
N(\omega,b) = \frac{\alpha Z^2}{\pi^2 \omega}\left| \int_0^{+\infty}  dk_{\perp} k_{\perp}^2 \frac{F\left((\frac{\omega}{\gamma})^2 + \vec{b}^2\right)}{(\frac{\omega}{\gamma})^2 + \vec{b}^2} \cdot J_1(b k_{\perp}) \right|^2 \,\,,
\end{eqnarray}
where { $\alpha$ is electromagnetic coupling constant, $\gamma$ is the Lorentz factor and $J_1$  is the Bessel function of the first kind}. For the nucleus, we will assume the  realistic form factor \cite{Klein:1999qj}, which corresponds to the Woods - Saxon distribution \cite{Woods:1954zz} and is the Fourier transform of the charge density of the nucleus, constrained by the experimental data. { Such a model provides a quite good  description of the form factors from heavy nuclei.}  In contrast, for a proton, we will derive the equivalent photon spectrum  from its elastic form factor in the dipole approximation (See e.g. \cite{david}). Finally, the  cross-section for the subprocess $\gamma \gamma \rightarrow \eta_t$ is given by
\begin{equation} \label{eq:epa}
\hat{\sigma}_{\gamma \gamma \rightarrow {\eta_t}}(\omega_{1}, \omega_{2}) = \int dW\ \delta(4\omega_{1}\omega_{2} - W) \frac{8\pi^2}{M_{\eta_t}} \Gamma_{{\eta_t} \rightarrow \gamma \gamma}(M_{\eta_t}) \delta(W - M_{\eta_t}^2),
\end{equation}
where $\Gamma_{{\eta_t} \rightarrow \gamma \gamma}$ is the partial decay width of the toponium into two photons. In our analysis, we will consider the recent values derived in Refs. \cite{Wang:2024hzd,Jiang:2024fyw}.

On the other hand, in order to estimate the exclusive toponium production by gluon - induced interactions in hadronic collisions, represented in the right panel of Fig. \ref{fig:diagram},  we consider the model proposed  two decades ago by Khoze, Martin and Ryskin~\cite{Khoze:2000cy, Khoze:2001xm, Khoze:2000jm}, denoted KMR model hereafter, which has been used to estimate different final states and have  predictions in reasonable agreement with the observed rates for exclusive processes at the Tevatron and LHC  (For a  review see Ref. \cite{review_lang}). In this model, { an additional $t$ channel gluon (denoted spectator gluon hereafter) is needed to screen the color and the cross - section for the $gg \rightarrow \eta_t$ subprocess is calculated considering a color singlet production. 
As a consequence, the incoming $gg$ state satisfies special selection rules, namely it has $J_z = 0$,
and positive $C$ and $P$ parity, which implies that only a subset of resonant $t\bar{t}$ states can be produced. As demonstrated in Refs. ~\cite{Khoze:2000cy, Khoze:2001xm, Khoze:2000jm},}
the total cross-section for the central exclusive production  of a toponium in hadronic collisions can be expressed in a factorized way, as follows
\begin{equation}
 \sigma(h_1 h_2 \xrightarrow{gg} h_1 \otimes {\eta_t} \otimes h_2) = \int dY \langle {\mathcal S}^2_{eik} \rangle {\mathcal L}_{excl} \frac{2\pi^2}{M_{\eta_t}^3} \Gamma(\eta_t \rightarrow gg)\,\,, 
\label{eq:kmr}
\end{equation}
where $\langle {\mathcal S}^2_{eik} \rangle$ is the gap survival probability (see below),  $\Gamma$ stand for the partial decay width of the toponium $\eta_t$ in a pair of gluons   and ${\mathcal L}_{excl}$ is the effective gluon - gluon luminosity, given by
\begin{equation}
 {\mathcal L}_{excl} = \left[ {\cal{C}} \int \frac{dQ_t^2}{Q_t^4} f_g(x_1,x_1^{\prime},Q_t^2, \mu^2) f_g(x_2,x_2^{\prime},Q_t^2, \mu^2)\right]^2\,\,.
\end{equation}
In this expression,  ${\cal{C}} = \pi/[(N_c^2 - 1)b]$, with $b$ the $t$-slope ($b = 4$ GeV$^{-2}$ in what follows \cite{Khoze:2001xm}), $Q_t^2$ is the virtuality of the soft gluon needed for color  screening { (the spectator gluon)}, $x_1$ and $x_2$ are  the longitudinal momentum fractions of the gluons which participate in the hard subprocess  and $x_1^{\prime}$ and $x_2^{\prime}$ the longitudinal momentum { fractions} of the spectator gluon. Moreover, the quantities $f_g$ are the  skewed unintegrated gluon densities. At leading logarithmic approximation, it is possible to express $f_g(x,x^{\prime},Q_t^2, \mu^2)$ in terms of the conventional integral gluon density and a Sudakov factor, which ensures that the active gluons that participate in the hard process do not radiate in the evolution from $Q_t$ up to the hard scale $\mu = m_{\perp} \equiv  \sqrt{ M_{\eta_t}^2 + p_{\eta_t,\perp}^2}$ (For details see Refs. \cite{Khoze:2000cy, Khoze:2001xm, Khoze:2000jm}). As the KMR model is implemented in the publicly available SuperChic Monte Carlo (MC) \cite{Harland-Lang:2020veo}, we  have modified this generator in order to include the possibility of estimating the $\eta_t$ production. Such a modification allow us to perform a full MC simulation of $\eta_t$ production in central exclusive processes. 
In this paper, we will calculate  $f_g$ using the CT18NNLO parametrization \cite{Hou:2019efy},   but we have verified that the impact of a different set on our results is small. Such a result is expected since in the case of toponium production we are probing large values of $x$ { ($\approx m_{\perp}/\sqrt{s} \gtrsim 10^{-2}$ for central rapidities)}, where the existing data provide important constraints in the behaviour of the  parton distribution functions (PDFs), which imply that the results obtained by different groups are similar. In addition, a similar result is obtained if other eigenvectors of the CT18 PDF are considered and the corresponding uncertainty is estimated. { In particular, the difference between the predictions is smaller than 5\%.} Such a result is also expected, { since
for $x \approx 10^{-2}$ the PDF uncertainty is constrained by already existing
experimental data  (See Fig. 2 (upper right panel)  in Ref. \cite{Hou:2019efy})}.  

We will estimate $\langle {\mathcal S}^2_{eik} \rangle$  using the model proposed in Ref. \cite{khoze13}, where the absorptive corrections associated to the additional soft hadron -- hadron interactions are estimated considering distinct approaches for the description of the diffractive data.   In particular, we will use the model 4 from Ref. \cite{khoze13}. { We have verified that if a different model for $\langle {\mathcal S}^2_{eik} \rangle$ is assumed, the predictions are reduced by a factor $\le 1.5$. As a consequence, the KMR predictions for the toponium production presented in this paper can be considered an upper limit.} Finally,  we will assume the values for the  partial decay width of the toponium into two gluons estimated in Refs. \cite{Wang:2024hzd,Jiang:2024fyw}.

\begin{table}[t]
\centering
\begin{tabular}{|c|c|c|}
\hline
 Channel & {\textbf{LHC}} & {\textbf{FCC}} \\
 \hline
 \hline
 $\gamma \gamma$ & 1.5 -- 13.2 & 11.6 --89.9 \\
 \hline
 $gg$ (KMR)  &   0.0028 -- 0.021        &  0.024 -- 0.19                  \\
 \hline
\end{tabular}
\caption{Cross - sections, in attobarn, for the { exclusive} toponium production by  photon  and gluon induced interactions in $pp$ collisions  at the LHC and FCC.  { The lower and higher values of the predictions are associated with the results for $\Gamma_{{\eta_t} \rightarrow \gamma \gamma}$ and $\Gamma_{{\eta_t} \rightarrow gg}$ derived in Refs. \cite{Wang:2024hzd} and \cite{Jiang:2024fyw}, respectively.}  }
\label{tab:pp}
\end{table}

\section{Results}
\label{sec:res}

In what follows, we will present our predictions for the exclusive toponium production in hadronic collisions. { One of the main inputs in our calculations are the values of the decay widths of the toponium into two photons or two gluons, which can be expressed in terms of the toponium wave function. In recent years, several groups have estimated such a quantity considering different frameworks and approximations \cite{Hagiwara:2008df,Sumino:2010bv,Aguilar-Saavedra:2024mnm,Fuks:2024yjj,Wang:2024hzd,Fu:2024bki,Jiang:2024fyw}, and derived distinct values for the mass of the $\eta_t$ state, as well as for the corresponding $\Gamma_{{\eta_t} \rightarrow \gamma \gamma}$ and $\Gamma_{{\eta_t} \rightarrow gg}$ decay widths. Here,  }  we will assume two different set of  values  for the $\eta_t$ mass  and decay widths: (a) $M_{\eta_t} = 343.62$ GeV, $\Gamma_{{\eta_t} \rightarrow \gamma \gamma} = 7.56$ keV and $\Gamma_{{\eta_t} \rightarrow gg} = 1.69$ MeV \cite{Wang:2024hzd}, and (b) $M_{\eta_t} = 341.267$ GeV,  $\Gamma_{{\eta_t} \rightarrow \gamma \gamma} = 57.18$ keV and $\Gamma_{{\eta_t} \rightarrow gg} = 12.96$ MeV  \cite{Jiang:2024fyw}. { These two sets provide, respectively, the lower and upper bounds for our results, and the large difference between them is mainly associated with the distinct treatment of the long - range part of the potential and the different assumptions for the fixing of the coefficient for the Coulomb part. It is important to emphasize that the description of the decay widths is the main source of uncertainty in our calculations. {In particular, Ref. \cite{Wang:2024hzd} also provides results for the decay widths considering a Coulomb potential, but its values are smaller than those derived in Ref. \cite{Jiang:2024fyw}, implying predictions that are between our lower and upper bounds. }

\begin{figure}[t]
	\centering
 \includegraphics[width=0.7\textwidth]{dsdW_toponium.eps} 
\caption{Invariant mass distribution for the exclusive toponium production by $\gamma \gamma$ interactions in $pp$ collisions at the FCC energy ($\sqrt{s} = 100$ TeV), obtained using  the partial decay width of the toponium into two photons derived in Refs. \cite{Wang:2024hzd} and \cite{Jiang:2024fyw}. For comparison, the prediction for the exclusive open $t\bar{t}$ production is also presented.}  
\label{fig:invmass}
\end{figure}

Initially, { we will}  consider $pp$ collisions at the LHC ($\sqrt{s} = 14$ TeV) and FCC ($\sqrt{s} = 100$ TeV)  energies. The  predictions for the total cross - sections are presented in Table \ref{tab:pp}. We have that the photon - induced interactions dominate the exclusive toponium production,  with the associated predictions being two orders of magnitude larger than the KMR one. Such a result is, in principle, expected, since in $pp$ collisions the effective photon - photon luminosity for producing a system of mass $M$ becomes larger than the gluon - gluon luminosity at large $M$, as demonstrated in Ref. \cite{Khoze:2001xm}.   In addition, our results indicate that the cross - sections increase by one order of magnitude at the FCC. However, the values are still of the order of few dozens of attobarn. Considering the expected integrated luminosities per year for the LHC (${\cal{L}}_{int}^{pp} \approx 160$ fb$^{-1}$) \cite{Bruning:2024zai}, we predict a maximum of 2 events. In contrast, this number of events becomes of the order of 90 at the FCC, where we expect to have  ${\cal{L}}_{int}^{pp} \approx 1$ ab$^{-1}$ \cite{FCC:2018vvp}. {Motivated by this result, we present in Fig. \ref{fig:invmass} the invariant mass distribution for the exclusive toponium production by $\gamma \gamma$ interactions in $pp$ collisions at the FCC energy ($\sqrt{s} = 100$ TeV), obtained using  the partial decay width of the toponium into two photons derived in Refs. \cite{Wang:2024hzd} and \cite{Jiang:2024fyw}. For comparison, the prediction for the exclusive open $t\bar{t}$ production, { computed for a stable top - quark}, is also presented. As expected for a resonance, the $\eta_t$ production implies a peak in the invariant mass  distribution at the corresponding mass, while the open $t\bar{t}$ production has a continuum distribution. Another important aspect, is that the peak is predicted to occur below the $2m_t$ threshold. Such aspect can be used, in principle, to separate the events associated with the resonance production from the continuum. The studies of exclusive processes in $pp$ collisions, performed by the ATLAS and CMS Collaborations using their forward detectors, have demonstrated that the mass of the central system can be reconstructed with a high precision  if the forward protons are tagged, which allow separating the events associated with a small range of invariant masses (See, e.g., Refs. \cite{Fiedler:2024mtl,Pitt:2024ovt}).  However, it is important to emphasize that the top decay can provide a contribution below threshold { already in case of the continuum}, and such possibility must be taken into account in order to separate the toponium from the continuum. Such aspect will be investigated in a future publication.

}

\begin{figure}[t]
	\centering
	\begin{tabular}{ccc}
    \\\includegraphics[width=0.5\textwidth]{toponium_LHC2.eps} & \, &
    \includegraphics[width=0.5\textwidth]{toponium_FCC2.eps}     
			\end{tabular}
\caption{Rapidity distributions for the exclusive toponium production by $\gamma \gamma$ interactions in $pp$, $pPb$ and $PbPb$ collisions at the LHC (left panel) and FCC (right panel). { The lower and upper values of the bands are associated with the results for the partial decay width of the toponium into two photons derived in Refs. \cite{Wang:2024hzd} and \cite{Jiang:2024fyw}, respectively.}  }
\label{fig:distribuicoes}
\end{figure}

For the nuclear case, previous studies for exclusive gluon - induced processes have pointed out that they are strongly suppressed by the additional soft hadron - hadron   interactions \cite{Levin:2008gi,Goncalves:2010dw,Basso:2017mue}, { that imply a very small value for $\langle {\mathcal S}^2_{eik} \rangle$,} with the production of large mass states predicted by the KMR model being much smaller that the results derived considering $\gamma \gamma$ interactions, which are enhanced by a factor $Z^2$ ($Z^4$) in 
 $pPb$ ($PbPb$) collisions (See, e.g. Refs. \cite{Goncalves:2015oua,Coelho:2020syp}).

 Based on these previous results and the fact that $\gamma \gamma$ channel already is dominant in $pp$ collisions, in what follows we only will  present the corresponding  predictions for the exclusive toponium production by photon - induced interactions. For $pPb$ ($PbPb$) collisions, we will estimate the  cross - sections for $\sqrt{s} = 8.16$ (5.5) TeV at the LHC and $\sqrt{s} = 63$ (39) TeV at the FCC. The results for the rapidity distributions of the toponium in $pp$, $pPb$ and $PbPb$ collisions are presented in Fig. \ref{fig:distribuicoes}, where the bands represent the current uncertainty in the value of $\Gamma_{{\eta_t} \rightarrow \gamma \gamma}$. Our results clearly demonstrate the increasing of the cross - section with the center - of - mass energy and that the distribution is asymmetric for $pPb$ collisions, since the photon fluxes associated with the proton and nucleus are distinct.  As expected, the predictions are largely enhanced in comparison to the $pp$ one. The corresponding results for the total cross - sections in $pPb$ and $PbPb$  collisions are presented in Table \ref{tab:nuclear}. 
 In particular, for $PbPb$ collisions, the cross - sections for the LHC (FCC) energy are four (six) orders of magnitude   larger than for $pp$ collisions, with the enhancement being smaller by a factor $\approx 10^2$ in the case of $pPb$ collisions.  Unfortunately, the increasing in the cross - sections due to the photon - flux enhancement does not imply a larger number of events in comparison to the results presented above, since the expected integrated luminosities for nuclear collisions are orders of magnitude smaller  than the value  for $pp$ collisions.  For example, at the FCC, the expected values are ${\cal{L}}_{int}^{pPb} \approx 8$ pb$^{-1}$ and ${\cal{L}}_{int}^{PbPb} \approx 33$ nb$^{-1}$ \cite{dEnterria:2022sut}. As a consequence, we predict a maximum of five events in these collisions at the FCC. Such results indicate that, given the current expected luminosities, an experimental analysis of the exclusive toponium production in these collisions is not feasible.

\begin{table}[t]
\centering
\begin{tabular}{|c|c|c|}
\hline
  & {\textbf{LHC}} & {\textbf{FCC}} \\
 \hline
 \hline
 $pPb$ & 518.2 -- 4083.5 & 21000 -- 165000 \\
 \hline
 $PbPb$  &   5071.1 -- 40697.6        &  $1.8\times 10^{7}$ -- $13.7\times 10^{7}$                  \\
 \hline
\end{tabular}
\caption{Cross - sections, in attobarn, for the toponium production by  photon induced interactions in $pPb$ and $PbPb$ collisions  at the LHC and FCC. { The lower and higher values of the predictions are associated with the results for $\Gamma_{{\eta_t} \rightarrow \gamma \gamma}$ derived in Refs. \cite{Wang:2024hzd} and \cite{Jiang:2024fyw}, respectively.} }
\label{tab:nuclear}
\end{table} 



\section{Summary}
\label{sec:sum}
 In this paper, we have performed an exploratory study of the exclusive toponium production in hadronic collisions, motivated by the recent results performed in inclusive processes that indicate a possible formation of a pseudoscalar $t\bar{t}$ state. One has considered photon and gluon - induced processes and demonstrated that the exclusive $\eta_t$ production is dominated by $\gamma \gamma$ interactions. We have estimated the rapidity distributions and total cross - sections for $pp$, $pPb$ and $PbPb$ collisions at the LHC and FCC energies.
Moreover, the corresponding number of events was calculated. Our results indicate that the study of the exclusive $\eta_t$ production in $pPb$ and $PbPb$ collisions is not feasible  considering the current expected integrated luminosities. However, the predictions for $pp$ collisions at the FCC motivate a more detailed analysis, including the $\eta_t$ decay and realistic experimental cuts, as well as the treatment of potential backgrounds associated. {In particular, the analysis of a toponium decaying into the constituent top and antitop quarks deserves a special attention due to the interplay between the continuous $t\bar{t}$ and the toponium production mechanisms, which makes the experimental separation a hard task. An alternative is to consider other toponium decay channels as e.g. into two gluons, which will generate two jets in the final state, or into two photons. If the experimental analysis focus on the $\eta_t \rightarrow \gamma \gamma$ decay channel,  the final state will be characterized by two rapidity gaps and the diphoton system, with the main background being} the exclusive diphoton production through the light - by - light scattering \cite{Goncalves:2020vvw,Coelho:2020syp}. In principle, such background can be strongly suppressed by considering the same strategy used to search for axionlike particles in exclusive processes \cite{Baldenegro:2018hng,Baldenegro:2019whq,Coelho:2020saz}. In these studies, the events are selected assuming a cut on the invariant mass of the final state in a range around the mass of the resonance,  since the presence of the resonance implies a peak  in the invariant mass distribution that dominates over the continuum associated with the { diphoton} background. Such aspect will be explored in a forthcoming study, where we intend to determine the observability (or not) of the toponium in exclusive processes at the FCC.

\begin{acknowledgments}
{ The authors are grateful to the anonymous referee for comments on a previous version
of the manuscript that helped significantly improve this
work.} V. P. G. would like to thank the members of the Institute of Nuclear Physics Polish Academy of Sciences for their warm hospitality during the completion of this
study. R. F. acknowledges support from the Conselho Nacional de Desenvolvimento Cient\'{\i}fico e Tecnol\'ogico (CNPq, Brazil), Grant No. 161770/2022-3. V.P.G. was partially supported by CNPq, FAPERGS and INCT-FNA (Process No. 464898/2014-5). D. E. Martins  acknowledges the support of POLONEZ BIS project No. 2021/43/P/ST2/02279
co-funded by the National Science Centre and the European Union’s Horizon 2020 research and innovation programme under the Marie Sklodowska-Curie Actions grant agreement no. 945339

\end{acknowledgments}


\begin{thebibliography}{}

\bibitem{Brambilla:2010cs}
N.~Brambilla, S.~Eidelman, B.~K.~Heltsley, R.~Vogt, G.~T.~Bodwin, E.~Eichten, A.~D.~Frawley, A.~B.~Meyer, R.~E.~Mitchell and V.~Papadimitriou, \textit{et al.}
Eur. Phys. J. C \textbf{71}, 1534 (2011)


\bibitem{Fadin:1990wx}
V.~S.~Fadin, V.~A.~Khoze and T.~Sjostrand,
Z. Phys. C \textbf{48}, 613-622 (1990)

\bibitem{Fabiano:1993vx}
N.~Fabiano, A.~Grau and G.~Pancheri,
Phys. Rev. D \textbf{50}, 3173-3175 (1994)

\bibitem{Hagiwara:2008df}
K.~Hagiwara, Y.~Sumino and H.~Yokoya,
Phys. Lett. B \textbf{666}, 71-76 (2008)

\bibitem{Sumino:2010bv}
Y.~Sumino and H.~Yokoya,
JHEP \textbf{09}, 034 (2010)
[erratum: JHEP \textbf{06}, 037 (2016)]



\bibitem{Fuks:2021xje}
B.~Fuks, K.~Hagiwara, K.~Ma and Y.~J.~Zheng,
Phys. Rev. D \textbf{104}, no.3, 034023 (2021)



\bibitem{Fadin:1987wz}
V.~S.~Fadin and V.~A.~Khoze,
JETP Lett. \textbf{46}, 525-529 (1987)

\bibitem{ATLAS:2019hau}
G.~Aad \textit{et al.} [ATLAS],
Eur. Phys. J. C \textbf{80}, no.6, 528 (2020)



\bibitem{ATLAS:2023gsl}
G.~Aad \textit{et al.} [ATLAS],
JHEP \textbf{07}, 141 (2023)




\bibitem{CMS:2024ybg}
A.~Tumasyan \textit{et al.} [CMS],
JHEP \textbf{02}, 064 (2025)




\bibitem{Aguilar-Saavedra:2024mnm}
J.~A.~Aguilar-Saavedra,
Phys. Rev. D \textbf{110}, no.5, 054032 (2024)

\bibitem{Fuks:2024yjj}
B.~Fuks, K.~Hagiwara, K.~Ma and Y.~J.~Zheng,
Eur. Phys. J. C \textbf{85}, no.2, 157 (2025)



\bibitem{Garzelli:2024uhe}
M.~V.~Garzelli, G.~Limatola, S.~O.~Moch, M.~Steinhauser and O.~Zenaiev,
Phys. Lett. B \textbf{866}, 139532 (2025)




\bibitem{Wang:2024hzd}
G.~L.~Wang, T.~F.~Feng and Y.~Q.~Wang,
Phys. Rev. D \textbf{111}, no.9, 096016 (2025)


\bibitem{Fu:2024bki}
J.~H.~Fu, Y.~J.~Li, H.~M.~Yang, Y.~B.~Li, Y.~J.~Zhang and C.~P.~Shen,
Phys. Rev. D \textbf{111}, no.11, 114020 (2025)

\bibitem{Jiang:2024fyw}
S.~J.~Jiang, B.~Q.~Li, G.~Z.~Xu and K.~Y.~Liu,
[arXiv:2412.18527 [hep-ph]].

\bibitem{CMS:2025kzt}
A.~Hayrapetyan \textit{et al.} [CMS],
[arXiv:2503.22382 [hep-ex]].



\bibitem{upc}
C. A. Bertulani and G. Baur, { Phys. Rep.} {\bf 163}, 299 (1988); F.~Krauss, M.~Greiner and G.~Soff,
  Prog.\ Part.\ Nucl.\ Phys.\  {\bf 39}, 503 (1997);
   C.~A. Bertulani, S.~R.~Klein and J.~Nystrand, Ann. Rev. Nucl. Part. Sci. {\bf 55}, 
271 (2005); V.~P.~Goncalves and M.~V.~T.~Machado,
  J.\ Phys.\ G {\bf 32}, 295 (2006);       A.~J.~Baltz {\it et al.},
  Phys.\ Rept.\  {\bf 458}, 1 (2008);       J.~G.~Contreras and J.~D.~Tapia Takaki,
  Int.\ J.\ Mod.\ Phys.\ A {\bf 30}, 1542012 (2015); 
      K.~Akiba {\it et al.} [LHC Forward Physics Working Group],
  J.\ Phys.\ G {\bf 43}, 110201 (2016); S.~R.~Klein and H.~Mantysaari,
Nature Rev. Phys. \textbf{1}, no.11, 662-674 (2019); S.~Klein and P.~Steinberg,
Ann. Rev. Nucl. Part. Sci. \textbf{70}, 323-354 (2020).

  


\bibitem{review_lang} 
  L.~A.~Harland-Lang, V.~A.~Khoze, M.~G.~Ryskin and W.~J.~Stirling,
 Int. Jour. Mod. Phys. A {\bf 29}, 1430031 (2014)

\bibitem{Goncalves:2020saa}
V.~P.~Gon\c{c}alves, D.~E.~Martins, M.~S.~Rangel and M.~Tasevsky,
Phys. Rev. D \textbf{102}, no.7, 074014 (2020)

\bibitem{Martins:2022dfg}
D.~E.~Martins, M.~Tasevsky and V.~P.~Goncalves,
Phys. Rev. D \textbf{105}, no.11, 114002 (2022)


\bibitem{Khoze:2000cy}
  V.~A.~Khoze, A.~D.~Martin and M.~G.~Ryskin,
  Eur.\ Phys.\ J.\  C {\bf 14}, 525 (2000) 

\bibitem{Khoze:2001xm}
  V.~A.~Khoze, A.~D.~Martin and M.~G.~Ryskin,
  Eur.\ Phys.\ J.\  C {\bf 23}, 311 (2002)

\bibitem{Khoze:2000jm}
  V.~A.~Khoze, A.~D.~Martin and M.~G.~Ryskin,
  Eur.\ Phys.\ J.\  C {\bf 19}, 477 (2001)
  [Erratum-ibid.\  C {\bf 20}, 599 (2001)]

\bibitem{Adamczyk:2015cjy}
  L.~Adamczyk {\it et al.},
  CERN-LHCC-2015-009, ATLAS-TDR-024.

\bibitem{Tasevsky:2015xya}
  M.~Tasevsky [ATLAS Collaboration],
  AIP Conf.\ Proc.\  {\bf 1654},    090001 (2015).
  
\bibitem{Albrow:2014lrm}
  M.~Albrow {\it et al.} [CMS and TOTEM Collaborations],
  CERN-LHCC-2014-021, TOTEM-TDR-003, CMS-TDR-13.


\bibitem{FCC:2018vvp}
A.~Abada \textit{et al.} [FCC],
Eur. Phys. J. ST \textbf{228}, no.4, 755-1107 (2019)






\bibitem{baur_jpg}
G.~Baur, K.~Hencken and D.~Trautmann,
  J.\ Phys.\ G {\bf 24}, 1657 (1998)



\bibitem{baur_ferreira}
  G.~Baur and L.~G.~Ferreira Filho,
  Nucl.\ Phys.\  A {\bf 518}, 786 (1990).


\bibitem{Azevedo:2019fyz}
C.~Azevedo, V.~P.~Gon\c{c}alves and B.~D.~Moreira,
Eur. Phys. J. C \textbf{79}, no.5, 432 (2019)


\bibitem{Klein:1999qj}
S.~Klein and J.~Nystrand,
Phys. Rev. C \textbf{60}, 014903 (1999)

\bibitem{Woods:1954zz}
R.~D.~Woods and D.~S.~Saxon,
Phys. Rev. \textbf{95}, 577-578 (1954)




\bibitem{david}
  D.~d'Enterria and J.~P.~Lansberg,
  Phys.\ Rev.\  D {\bf 81}, 014004 (2010)





\bibitem{harland} 
  L.~A.~Harland-Lang,
  Phys.\ Rev.\ D {\bf 88},  034029 (2013)
  
  \bibitem{khoze13} 
  V.~A.~Khoze, A.~D.~Martin and M.~G.~Ryskin,
  Eur.\ Phys.\ J.\ C {\bf 73}, 2503 (2013)

\bibitem{Hou:2019efy}
T.~J.~Hou, J.~Gao, T.~J.~Hobbs, K.~Xie, S.~Dulat, M.~Guzzi, J.~Huston, P.~Nadolsky, J.~Pumplin and C.~Schmidt, \textit{et al.}
Phys. Rev. D \textbf{103}, no.1, 014013 (2021)




\bibitem{Harland-Lang:2020veo}
L.~A.~Harland-Lang, M.~Tasevsky, V.~A.~Khoze and M.~G.~Ryskin,
Eur. Phys. J. C \textbf{80} (2020) no.10, 925


\bibitem{Levin:2008gi}
E.~Levin and J.~Miller,
[arXiv:0801.3593 [hep-ph]].



\bibitem{Goncalves:2010dw}
V.~P.~Goncalves and W.~K.~Sauter,
Phys. Rev. D \textbf{82}, 056009 (2010)

\bibitem{Basso:2017mue}
E.~Basso, V.~P.~Goncalves, A.~K.~Kohara and M.~S.~Rangel,
Eur. Phys. J. C \textbf{77}, no.9, 600 (2017)

\bibitem{Goncalves:2015oua}
V.~P.~Goncalves and W.~K.~Sauter,
Phys. Rev. D \textbf{91}, 035004 (2015)


\bibitem{Coelho:2020syp}
R.~O.~Coelho, V.~P.~Gon\c{c}alves, D.~E.~Martins and M.~Rangel,
Eur. Phys. J. C \textbf{80}, no.5, 488 (2020)

\bibitem{Bruning:2024zai}
O.~Br\"uning and L.~Rossi,
``The High Luminosity Large Hadron Collider,''
World Scientific, 2024,
ISBN 978-981-12-7894-5
doi:10.1142/13487

  
\bibitem{Fiedler:2024mtl}
P.~Fiedler [ATLAS Forward Detectors],
PoS \textbf{ICHEP2024}, 875 (2025)
  
\bibitem{Pitt:2024ovt}
M.~Pitt [CMS],
PoS \textbf{LHCP2023}, 012 (2024)

  
\bibitem{dEnterria:2022sut}
D.~d'Enterria, M.~Drewes, A.~Giammanco, J.~Hajer, E.~Bratkovskaya, R.~Bruce, N.~Burmasov, M.~Dyndal, O.~Gould and I.~Grabowska-Bold, \textit{et al.}
J. Phys. G \textbf{50}, no.5, 050501 (2023).


\bibitem{Goncalves:2020vvw}
V.~P.~Goncalves, D.~E.~Martins and M.~S.~Rangel,
Eur. Phys. J. C \textbf{80}, no.9, 841 (2020)


\bibitem{Baldenegro:2018hng}
C.~Baldenegro, S.~Fichet, G.~von Gersdorff and C.~Royon,
JHEP \textbf{06}, 131 (2018)

\bibitem{Baldenegro:2019whq}
C.~Baldenegro, S.~Hassani, C.~Royon and L.~Schoeffel,
Phys. Lett. B \textbf{795}, 339-345 (2019)



\bibitem{Coelho:2020saz}
R.~O.~Coelho, V.~P.~Goncalves, D.~E.~Martins and M.~S.~Rangel,
Phys. Lett. B \textbf{806}, 135512 (2020)


\end{thebibliography}
\end{document}